\documentclass[aps,twocolumn,prl,superscriptaddress,nofootinbib,floatfix,longbibliography]{revtex4-2}

\usepackage{dcolumn}
\usepackage{bm}
\usepackage{amssymb}
\usepackage{amsfonts}
\usepackage{amsmath}
\usepackage{graphicx}
\usepackage{epstopdf}
\usepackage{float}
\usepackage{relsize}
\usepackage{mathtools}
\usepackage{placeins}
\usepackage{array}
\usepackage{xcolor}
\begin{document}

\title{Quantum field theory-inspired cure for black hole singularities}

\author{L. C. N. Santos}
\email{luis.santos@ufsc.br}
\affiliation{Departamento de F\'isica, CFM - Universidade Federal de Santa Catarina; C.P. 476, CEP 88.040-900, Florian\'opolis, SC, Brasil.}

\begin{abstract}
In recent years, there has been a growing interest in the study of regular black holes, driven by the search for singularity-free geometries. This research has revealed intriguing similarities between the regularization mechanisms used in black hole models and those employed in quantum field theory, such as the introduction of exponential suppression or energy cutoffs. We propose a systematic exponential cutoff regularization scheme for static, spherically symmetric black hole solutions in general relativity. The method explored in this paper serves as an alternative to the black-bounce singularity suppression mechanism proposed by Simpson and Visser, which involves a coordinate remapping $r \rightarrow\sqrt{r^2+a^2}$, as well as to the mechanism proposed by Bronnikov, which employs a Bardeen-type remapping in the metric. The method presented here introduces exponential factors in the mass function, smoothing curvature divergences and ensuring geodesic completeness under specific conditions. This approach allows the regularization of known singular spacetimes without altering their asymptotic structure. We analyze curvature invariants, horizon formation, and thermodynamic properties, showing that the regularized geometries avoid singularities while maintaining physical consistency. As examples of application, we regularize the Schwarzschild black hole and present a novel regularized Kiselev solution. The method provides a unified framework to systematically generate singularity-free black holes within classical general relativity
\end{abstract}


\maketitle

\preprint{}

\volumeyear{} \volumenumber{} \issuenumber{} \eid{identifier} \startpage{1} %
\endpage{}
\section{Introduction}
Penrose's 1965 singularity theorem \cite{Penrose:1964wq} establishes that, under certain physically reasonable conditions, the formation of singularities in spacetime is inevitable (See Refs.~\cite{Senovilla:1998oua,Lan:2023cvz} for a comprehensive discussion). Consequently, in order to obtain regular solutions that avoid such singularities, one must circumvent at least one of the assumptions underlying the theorem. It is not straightforward to determine how the theorem is circumvented in the case of regular black holes. In particular, although the theorem is usually formulated under an energy condition that ensures the formation of singularities, many known regular solutions may still satisfy the strong energy condition, for example, and yet correspond to regular black hole spacetimes \cite{Borissova:2025msp}. The key point here is that the evasion of the theorem in such cases must occur through the violation of some other assumption underlying the theorem. 

The issue of regularity in a spacetime can be assessed in at least two complementary ways. First, by analyzing curvature invariants such as the Kretschmann scalar or the Ricci scalar one can detect potential singularities through the divergence of these quantities. Second, regularity can be examined through the concept of geodesic completeness, which refers to whether all geodesics can be extended to arbitrary values of their affine parameter. A spacetime is said to be geodesically complete if no freely falling observer or light ray reaches a boundary of the manifold in finite proper time or affine parameter. Geodesic incompleteness is often taken as a hallmark of spacetime singularities in general relativity. There has been a growing effort to construct and study solutions of the field equations that yield regular geometries~\cite{Ditta:2025csz,Mustafa:2025cou,Anand:2025mlc,Waseem:2025bwb,Mannobova:2025uqf,Shahzad:2025msa,Vacaru:2025ngf,Calza:2025mrt,Mohamed:2025iqn,Song:2025qpo,Singh:2025ald,Alonso-Bardaji:2025qft,Kar:2025phe,Urmanov:2025nou,Mushtaq:2025ewk,Lin:2025zea,Lutfuoglu:2025hwh,Fernandes:2025eoc,Vertogradov:2025jxp,Capozziello:2025ycu,Ladghami:2025qjx,Huang:2025uhv,Khoo:2025qjc,Turakhonov:2025ojy,Coviello:2025pla,Barenboim:2025ckx,Ghaderi:2025xxw,Abu-Saleem:2025gfj,Harada:2025cwd,NunesdosSantos:2025alw,Casadio:2025pun,Javed:2025bpr,Konoplya:2025hgp,Borissova:2025msp,Narzilloev:2025nof,Khoshrangbaf:2025bwg,Chaudhary:2025lzi}. A particularly popular strategy to achieve this involves coupling the field equations with nonlinear electrodynamics~\cite{ayon1998bardeen,bronnikov2001regular,han2020thermodynamics,bronnikov2020comment,junior2024regular,walia2024exploring,dolan2024superradiant,tangphati2024magnetically,guo2024recovery,kar2024novel,bronnikov2024regular}. The emergence of regularity in black hole solutions may be closely related to quantum gravitational effects. In this context, it is possible to derive regular black holes as quantum
corrections to classical singular black holes~\cite{modesto2004disappearance,gambini2008black,ashtekar2023regular,modesto2006loop,momennia2022quasinormal,sharif2010quantum}. 

In physics, singularities mark points where a theory ceases to produce finite, well-defined predictions. In quantum field theory (QFT), ultraviolet divergences arise in perturbative calculations, but regularization methods such as the exponential cutoff can control high-energy contributions. Considering spherically symmetric solutions in general relativity that are regular, some of them introduce an exponential factor $e^{-a/r}$ coupled to the usual mass $m$ of a Schwarzschild black hole. Here, $a$ has units of length and represents a ``smoothing scale'' in real space. It is interesting to note the similarity between this class of regular geometry and the exponential cutoff regularization method where $a$ is analogous to the scale $\Lambda^{-1}$ that suppresses high-frequency modes in QFT with a cutoff term $e^{-p/\Lambda}$. Therefore, one can heuristically identify the cutoff scales in both frameworks as $a \sim \frac{1}{\Lambda}$. In addition, both replace idealized singular behaviors (UV divergence in QFT, curvature singularity in GR) with smooth, finite ones. 

In this work, we investigate a method that offers an alternative to the singularity avoidance approach introduced by Simpson and Visser \cite{Simpson:2018tsi}, based on the coordinate transformation \( r \rightarrow \sqrt{r^{2}+a^{2}} \), and also to the proposal by Bronnikov \cite{Bronnikov:2024izh}, which applies a Bardeen-inspired modification directly to the metric. We propose a systematic procedure to regularize static, spherically symmetric solutions in general relativity through a general exponential factor. 
We conduct a detailed analysis of solutions to Einstein's field equations associated with regular black holes, including the study of curvature invariants and geodesic completeness. Two regularization mechanisms are proposed, both involving exponential terms coupled to a mass function that ensures the regularity of the geometry. We apply the procedure to two cases: the spherically Schwarzschild black hole and a black hole surrounded by quintessence. In the latter case, the solution obtained corresponds, to our knowledge, to a new regular solution of Einstein's field equation.
 
Our goal here is not to merely generate new solutions to Einstein's field equations associated with regular black holes, but rather to regularize any existing solution that meets the criteria we have defined. As a result, there will be changes in the geometry near the black hole (which can be made more precise by introducing the scale parameter $a$) in the regularized cases. This approach yields a broad set of new geometries with distinct properties, which can be explored in detail at a later stage. 

\section{Formalism}
In this section we intend to introduce the systematic scheme to obtain regular solutions from known solutions that are singular at particular points of the spacetime. We restrict our analysis to metric of spacetimes whose undetermined functions depend only on the radial coordinate, written in the form 
\begin{equation}
    ds^2 = -B(r) dt^2 + A(r) dr^2 + r^2(d\theta^2 + \sin^2\theta \, d\varphi^2),
    \label{eq1}
\end{equation}
where 
\begin{equation}
B(r) = 1/A(r) = 1 - \frac{2M(r)}{r},
   \label{eq2} 
\end{equation}
and the curvature invariants are written as
\begin{align}
        R =& \frac{4 M'}{r^2} + \frac{2 M''}{r},\label{eq3}\\
        R_{\mu\nu} R^{\mu\nu} =& \frac{8 {M'}^2}{r^4} + \frac{2 {M''}^2}{r^2},  \label{eq4}\\
        R_{\mu\nu\lambda\rho} R^{\mu\nu\lambda\rho} =& \frac{48 M^2}{r^6} - \frac{16 M}{r^3} \left( \frac{4 M'}{r^2} - \frac{M''}{r} \right) \nonumber \\
        &+ 4 \left( \frac{8 {M'}^2}{r^4} - \frac{4 M' M''}{r^3} + \frac{{M''}^2}{r^2} \right).
        \label{eq5}
\end{align}
The regularity conditions in the context of curvature invariants imply that if the relation $M(r)/r^3$ is finite then all of the terms in Eqs.~(\ref{eq3}), (\ref{eq4}), and (\ref{eq5}) are finite in the limit $r \rightarrow 0$. An feature of using curvature invariants to assess spacetime regularity lies in their coordinate-independent nature. These scalar quantities, constructed from contractions of the Riemann tensor, provide a manifestly covariant criterion for detecting singular behavior. In contrast, the analysis of geodesic completeness typically requires solving the geodesic equations, which depend on the explicit choice of coordinates. 
We consider a specific functional form for the mass function \( M(r) \) appearing in the metric component \( B(r) = 1 - 2M(r)/r\), namely
\begin{equation}
M(r) = m(r)\, \exp\left(-\frac{a^{n+1}}{r^{n+1}}\right),\:\: n \in \mathbb{N},
\label{eq6}
\end{equation}
where the function \( m(r) \) should have subexponential growth. This implies that $m(r)$ exhibits a polynomial behavior, or even an exponential one if such an exponential function grows more slowly than the exponential in Eq.~(\ref{eq6}). As we will see, 
this class of functions covers most of the known solutions, such as, for example, the Schwarzschild solution for \( m(r) = m_0 = \text{constant} \), the Reissner–Nordström solution for \( m(r) = m_0 - Q^2/2r \), among others. The exponential factor acts as a regulating function that introduces a smooth cutoff near the origin. This construction provides a general prescription to regularize a wide class of static, spherically symmetric solutions by suppressing the divergent behavior of \( m(r) \) as \( r \to 0 \), while recovering the original solution asymptotically at large distances. In this way, we can see that the dominant term $M(r)/r^3=m(r)\exp(-a^{n+1}/r^{n+1})/r^3$ in the limit $r \rightarrow 0$ vanishes if $m(r)$ grows at most polynomially or its growth rate remains slower than that of the exponential in Eq. (\ref{eq6}). 

It is worth noting that the focus here is not to construct regular solutions but rather to remove the existence of singularities of known solutions in a way analogue to the regularization scheme in QFT that removes the existing infinities in the theory.

\subsection{General spacetime and source }
Let us consider here an anisotropic fluid that is associated with the energy-momentum tensor \cite{Herrera:1997plx,santos00}
\begin{equation}
    T_{\mu\nu} = p_t g_{\mu\nu} + (p_t + \rho) U_{\mu} U_{\nu} + (p_r - p_t) N_{\mu} N_{\nu},
    \label{eq7}
\end{equation}
where \( p_t(r) \), \( \rho(r) \), and \( p_r(r) \) denote the tangential pressure, the energy density, and the radial pressure of the fluid, respectively. The vectors \( U_{\mu} \) and \( N_{\mu} \), representing the four-velocity and the unit radial direction satisfying the normalization conditions \( U_{\nu} U^{\nu} = -1 \), \( N_{\nu} N^{\nu} = 1 \), and \( U_{\nu} N^{\nu} = 0 \).  In the case of the spacetime (\ref{eq1}), the components of the energy-momentum tensor are written as $ T^{\mu}_{\:\:\:\nu} = \text{diag}\left(-\rho, p_r, p_t, p_t\right)$. Considering the Einstein field equations, $G^{\mu}_{\:\:\:\nu} = 8\pi T^{\mu}_{\:\:\:\nu}$, the static and spherically symmetric line element defined in (\ref{eq1}), and  the components of the anisotropic energy-momentum tensor, the field equations take the form:
\begin{align}
   G^{t}_{\:\:t}=G^{r}_{\:\:r}=& \frac{1}{r}\frac{dB(r)}{dr} + \frac{B(r)}{r^2} - \frac{1}{r^2}= -8\pi\rho, \label{eq8}\\
 G^{\theta}_{\:\:\theta}=G^{\phi}_{\:\:\phi}=& \frac{1}{2}\frac{d^2B(r)}{dr^2}+\frac{1}{r}\frac{dB(r)}{dr} = 8\pi p_t
    \label{eq9}.
\end{align}
Equations (\ref{eq8}) and (\ref{eq9}) can be solved for $\rho$ and $p_t$, with the result expressed as a ratio between $\rho$
and $p_t$, effectively yielding an equation of state of the form $\rho = w(r)p_t$ with
\begin{equation}
    w(r) = - \frac{2 r^{2\bar{n}} \left(2 a^{2\bar{n}} \bar{n} m + r^{1 + 2\bar{n}} m' \right)}{
4 a^{4\bar{n}} \bar{n}^2 m - 
2 a^{2\bar{n}} \bar{n} r^{2\bar{n}} \bar{C}_{\bar{n}} + 
r^{2 + 4\bar{n}} m''}.
 \label{eq10}
\end{equation}
where $\bar{C}_{\bar{n}}=m + 2\bar{n} m - 2r m' $ and $\bar{n}=(n+1)/2$. Thus, several solutions for a particular form of $m(r)$ can be associated with this equation of state. The regularity of curvature invariants is guaranteed by the exponential in the mass function.

\subsection{A regularization scheme based on Dymnikova's solution}
A regular black hole with an exponential term was proposed by Dymnikova in the following form~\cite{dymnikova1992vacuum}: $B(r) = 1 - 2M \left(1 - e^{-r^3 / R^3}\right)/r$ where $R$ is associated with a characteristic scale of spacetime. From the point of view of our scheme, we can try the term $1 - e^{-r^3 / R^3}$ as a regularizing factor for a general mass function $m(r)$ in the spirit of Eq.~(\ref{eq6}). Although this term regularizes Schwarzschild geometry, for spacetimes involving a mass function in the form $m(r) \propto r^{-1},r^2,r^3,...$ we need to change the exponential term to the fourth order in $r$ in view of the regularity of the curvature invariants. The aforementioned analysis suggests the following mass function
\begin{equation}
M(r) = m(r)\left[1-\exp\left(-\frac{r^{n+3}}{R^{n+3}}\right)\right],\:\: n \in \mathbb{N},
\label{eq11}
\end{equation}
such that if the function $m(r)$ is of polynomial type or the term $M(r)/r^3$ remains regular near the origin, the invariants are regular. The usual Dymnikova solution is recovered for $n=0$ and $m(r)=m_0 =\text{constant}$, however, solutions for $n>0$ are also regular and have a geometric behavior distinct from the usual Dymnikova black hole. Thus, we see that for a Dymnikova-type regularization term, we must choose $n>1$ in order to regularize the known solutions. Due to the simplicity and comprehensiveness of the regularization scheme proposed in ($\ref{eq6}$), we will choose to focus on the analysis of this scheme in the rest of the paper.

\section{Geodesic completeness}
\begin{figure}[h]
\centering
\includegraphics[scale=0.62]{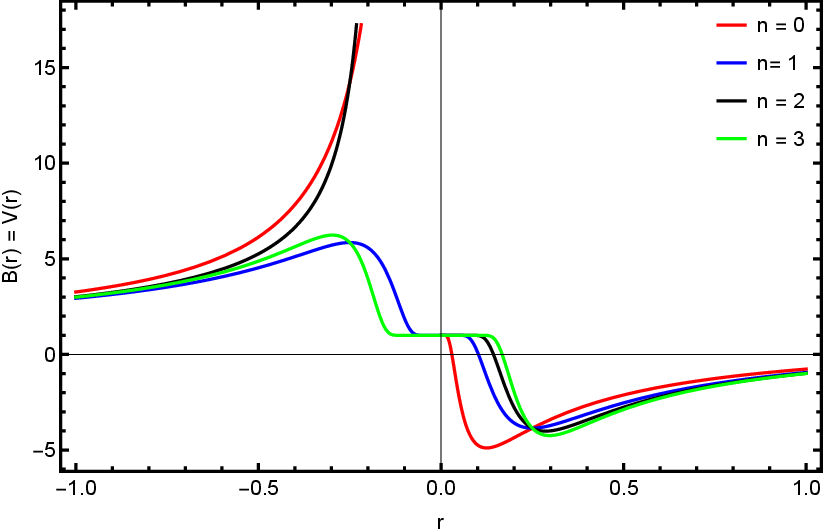} 
\caption{The plot shows a general  metric solution $B(r)$ with $m(r)$ exhibiting a subexponential growth as a function of the radial coordinate $r$ for different values of the parameter $n$, with fixed parameters $a = 1/4$ and $M = 1$.    
}
\label{fig1}
\end{figure}
In this section, we will establish the conditions under which the regularized solution is not only free of curvature singularities but also geodesically complete. A spacetime is considered \emph{geodesically complete} if all trajectories of freely falling observers and light rays can be extended indefinitely along their proper time or affine parameter, without ever encountering the edge of the manifold. When a spacetime contains geodesics that end after a finite proper time or affine parameter, it is referred to as \emph{geodesically incomplete}, a feature that is widely interpreted as a signature of singularities in general relativity. In the case of spacetime where the proper time required to reach $r=0$ is finite, it is mathematically possible to extend $r$ to negative values. In this case, the extended spacetime can be analyzed to determine whether singular points arise, or if the spacetime remains regular within this domain.

We start by representing the four-velocity as
$u^{\mu}=dx^{\mu}/d\tau=\Dot{x}^{\mu}.$
Owing to the symmetry of the system, the motion can be confined to the equatorial plane by setting $\theta =\text{constant} =\pi/2$, which does not influence the investigation of geodesics of massive particles. From definition of four-velocity, the inner product becomes:
\begin{equation} u^{\mu}u_{\mu}=-\kappa=-B(r)\Dot{t}^2+B(r)^{-1}\Dot{r}^2 + r^2\Dot{\varphi}^2. \label{3c2} \end{equation}
The parameter $\kappa=1$ denotes geodesics of time-like nature, whereas $\kappa=0$ identifies null geodesics. The geometry defined by the metric function admits two Killing vectors: $t^{\mu}=(1,0,0,0)$ and $\varphi^{\mu}=(0,0,0,1)$. These vectors correspond to two conserved physical quantities, the energy $E$ and angular momentum $L$ which are given by:
\begin{equation} -g_{\mu\nu}t^{\mu}u^{\nu}=B(r)\Dot{t}=E,\:\:\: g_{\mu\nu}\varphi^{\mu}u^{\nu}=r^2\Dot{\varphi}=L. \label{3c3} \end{equation}
Inserting these expressions into Eq.~(\ref{3c2}) yields a expression which can be rearranged as:
\begin{equation} \Dot{r}^2 + V(r)= E^2, \:\: V(r)=\frac{L^2 B(r)}{r^2}+\kappa B(r).
\label{3c5} \end{equation}
where $V(r)$ is an effective potential. For massive particles and purely radial motion, we choose a path where the Eq.~(\ref{3c5}) can be integrated so that the proper time is written as
\begin{equation}
    \tau = \int_{r}^{r_i}dr\frac{1}{\sqrt{E^2 - B(r)}}.
    \label{tau}
\end{equation}
It is straightforward to see that if $m(r)$ exhibits subexponential growth, $\textit{i.e.}$, there exists $\delta>0$ such that $m(r) \leq \exp{(\epsilon/r)}$ for all $0<r<\delta$ and $0<\epsilon<a^{n+1}$, then the function $M(r)$ defined in Eq.~(\ref{eq6}) tends to zero as $r \rightarrow 0^+$. If $E^2 <1$, the test particle does not have enough energy to overcome the potential barrier and will bounce back. For $E^2=1$, the proper time Eq.~($\ref{tau}$) to reach $r=0^+$ is infinity since the integrand behaves asymptotically as $\sim \exp(-a^{n+1}/2r^{n+1})$ for the choice $\epsilon=a^{n+1}/2$. If $E^2>1$, the particle has sufficient energy to overcome the potential barrier near $r=0^+$ and $E^2-B(r) \approx E^2 -1$ since $B(r)\rightarrow 1$ near $r=0^+$. This leads to a finite integral $\tau \approx \text{constant}$. Thus, based on the previous discussion, we conclude that the spacetime must be extended to the region $r < 0$. In this extended domain, one must consider that, depending on the behavior of the exponential function for certain values of $n$, the metric may fail to be continuous at $r = 0$, \textit{i.e.}, the left and
right limits of $B(r)$ are different at this point. 

To directly illustrate the behavior of the geometry in this region, we choose a function for $m(r)$ that grows similarly to $\sim\exp{(\epsilon/r)}$, but still exhibits a slower growth rate since we assume $\epsilon < a^{n+1}$. We can see in Fig.~\ref{fig1} that metric function $B(r)$ is continuous for $n=0,1,3,...$ but if $n$ is even, then the spacetime is not regular at $r=0.$ Continuing the analysis for odd values of $n$ in the negative domain of $r$, we conclude that the proper time required to reach the edges at $r = \pm \infty$ is infinite, indicating that the spacetime is geodesically complete.  
It is worth noting that the analysis conducted thus far is valid for any function $m(r)$ exhibiting subexponential growth.

\begin{equation}  \label{3c6} \end{equation}
\subsection{Horizon structure}
Considering the possible existence of horizons in the regularized solutions, we must analyze the roots of the equation $g_{tt} = 0$. Due to the presence of the exponential regularizing term, it is not always possible to obtain analytical solutions for a given mass function $m(r)$. Here, we will once again follow the strategy of studying the function $m(r)$ with a more pronounced growth, while still ensuring that the solution $B(r)$  remains regular. For this purpose, it is useful to consider a mass function that grows as fast as possible while still being subexponential, meaning that its growth rate remains slower than that of the exponential regularizing function. Naturally, if the solution is regular for this mass function, it will also be regular for polynomial mass functions, which includes a large class of static spherically symmetric solutions of the Einstein field equations. By setting $m(r)=m_0\exp(\epsilon^{n+1}/r^{n+1})$ in Eq.~(\ref{eq11}) with $\epsilon^{n+1}<a^{n+1}$, we  obtain the equation for the horizon $1-(2m_0 /r_h )e^{-\bar{a}/r_{h}^{n+1}}=0$ where $\bar{a}\equiv a^{n+1}-\epsilon^{n+1}$. This equation can be solved in terms of the Lambert $W(x)$ function ~\cite{Valluri:2000zz}, and expressed as:
\begin{equation}
    r_h = 2m_0 e^{\frac{W(x)}{n+1}},\:\:x \equiv -\frac{\bar{a}(n+1)}{(2m_0)^{n+1}}.
    \label{eq17}
\end{equation}
The Lambert $W$ function is defined as the solution to the equation $We^{W}=z$, where $z$ in the general case is a complex number. Possible solutions for $W_k(z)$ are labeled by $k=0,\pm1,\pm2,\text{etc}$. The branches associated with real solutions $W_{0}(x)$ and $W_{-1}(x)$ can be solved only if $x \geq -1/e$ in the case of $W_{0}(x)$ and $-1/e \leq x<0$ in the case of $W_{-1}(x)$. It follows that the range of the horizon described in (\ref{eq17}) is 
\begin{align}
     & r_h \in \left[ 2m_0 e^{-\frac{1}{n+1}},
      +\infty \right),\: \text{for}\:\: W_0(x), \label{eq18}\\
      & r_h \in \left( 0, 2m_0 e^{-\frac{1}{n+1}} \right],\: \text{for}\:\: W_{-1}(x).
      \label{eq19}
\end{align}
In light of this result, we conclude that the possible values for the horizon range from zero to infinity. We observe that $ W_{-1}(x)$ is associated with an inner horizon, while $W_0(x)$ corresponds to an outer horizon. On the other hand, we can obtain the surface gravity $\kappa$ defined by equation $\nabla_{\nu} \left( - \xi^{\mu} \xi_{\mu} \right) = 2 \kappa \xi_{\nu}$, where $\xi^{\mu} =(1,0,0,0)^{\mu}$ is the Killing vector. Thus, Hawking radiation $T=\hbar \kappa/2\pi k_B$ can be attributed to both the inner and outer horizons in the form
\begin{equation}
    T_{\pm}=\frac{\hbar m_0 r_{\pm}^{-2}}{2\pi k_{B}}\left(\frac{a}{r_{\pm}}\right)^{n+1}e^{-\left(\frac{a}{r_{\pm}}\right)^{n+1}}\left[ -n-1+\left(\frac{r_{\pm}}{a}\right)^{n+1}\right],
\end{equation}    
where $r_{\pm}=2m_0 e^{\frac{W_{\pm}(x)}{n+1}}$ in which the positive sign is associated with the radius of the outer horizon and the negative sign with the inner horizon, with $W_{+}(x) \equiv W_{0}(x)$ and $W_{-}(x) \equiv W_{-1}(x)$.
   

\section{Regularizing Schwarzschild and Kiselev solutions } 
Here, we present examples illustrating the application of the method developed thus far. As we will see, in the case of the Schwarzschild black hole, we explore specific values of 
$n$, some of which reproduce well-known geometries. For the Kiselev black hole, we regularize the geometry as an example and obtain a solution that, to the best of our knowledge, is new to the literature.
     \subsection{Schwarzschild black hole}
\begin{figure}[h]
\centering
\includegraphics[scale=0.62]{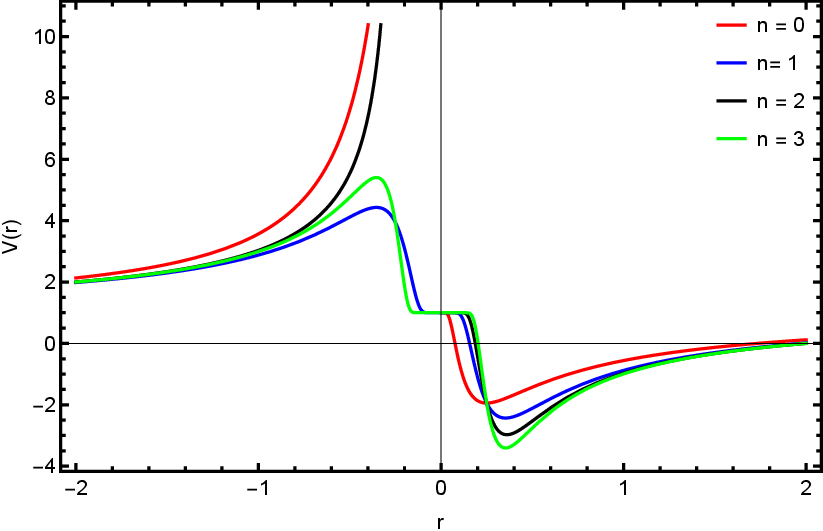} 
\caption{The plot shows the effective potential \( V(r) \) as a function of the radial coordinate \( r \) for different values of the parameter \( n \), with fixed parameters \( a = 1/4 \) and \( M = 1 \). As \( n \) increases from 0 to 3, the shape of the potential changes significantly. For negative values of \( r \), the potential becomes steeper and more repulsive with increasing \( n \). Near \( r = 0 \), a pronounced potential well appears, whose depth and profile vary depending on \( n \), indicating the presence of a strong interaction in this region.   
}
\label{fig2}
\end{figure}
We begin by considering the Schwarzschild solution, which in our discussion is associated with a mass function of the form $m= m_0$. Using the first value $n = 0$ in Eq.~(\ref{eq6}), we obtain the specific form of the function \( B(r) =1 - (2m_0/r)\exp{(-a/r)}\). On the other hand, the charged version of this solution, discussed in~\cite{Culetu:2014lca}, is given by mass function $m(r)= m_0-Q^2/2r$. The anisotropic fluid acts as source for the ``charged" solution. Indeed, for $r$ sufficiently large relative to $a$, the pressures and energy density of fluid resembles the Maxwell energy-momentum tensor. This black hole has interesting properties such as entropy given by simply $A/4$. Regarding the study of geodesic completeness of this geometry, we can see that the geometry has an asymptotic Minkowski core
where the curvature invariants vanish in the limit $r\rightarrow 0^+$. However, the  proper time for massive particles  to reach $r\rightarrow 0^+$ 
is finite. This imply that the spacetime should be extend beyond $r=0$ \cite{Zhou:2022yio}, but this geometry is ill defined at $r=0$ and consequently geodesically incomplete. 

For $n=1$ and $m(r)=m_0$, the solution is written as $B(r)=1-(2m/r)\text{e}^{-a^2/r^2}$ that is the geometry studied in \cite{Xiang:2013sza}. This spacetime represents a slight modification of the $n = 0$ case. It asymptotically approaches the Schwarzschild solution for large values of $r$ and exhibits a Minkowski core at the origin. It can be shown that massive particles reach $r = 0$ within a finite proper time, implying that the spacetime must be extended, as in the previous case. However, in contrast to the $n = 0$ case, the geometry is now regular at $r = 0$, making the extension to $r < 0$ not only possible but also well-defined \cite{Zhou:2022yio}. The analysis of the proper time required for a particle to reach the edges at $r = \pm \infty$ yields an infinite value, indicating that the spacetime is geodesically complete. From these examples of regularized geometries, we conclude that the condition for geodesic completeness follows from the choice of an even $n$ as discussed in the previous section. Fig.~\ref{fig2} illustrates the behavior of the potential for different values of 
$n$, where discontinuous curves can be identified. We are now in a position to address the regularization of more complex geometries using the present formalism.
     
\subsection{Novel regular Kiselev solution}
\begin{figure}[h]
\centering
\includegraphics[scale=0.62]{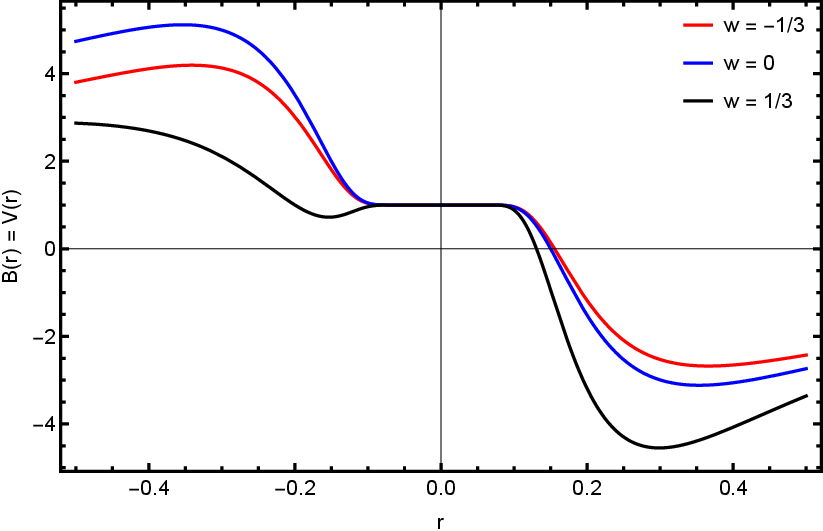} 
\caption{The plot shows a metric solution $B(r)$ with $m(r)$ being the Kiselv solution as a function of the radial coordinate $r$ for different values of the parameter $w$, with fixed parameters $a = 1/4$, $M = 1$ and $n=1$.    
}
\label{fig3}
\end{figure}

The Kiselev solution describes a static spherically symmetric black hole surrounded by a quintessential matter field. Unlike the standard Schwarzschild or Reissner–Nordström metrics, the Kiselev metric accounts for the presence of a fluid. In this solution, we have an equation of state parameter $\omega$,  effectively mimicking $p = \omega  \rho$, after integration over all angles, where $p$ is the pressure and $\rho$ the energy density with the metric function given by
\begin{equation}
f(r) = 1 - \frac{2m_0}{r} - \frac{c}{r^{3\omega + 1}},
\label{eq21}
\end{equation}
where $m_0$ is the mass of the black hole and $c$ is a constant related to the intensity of the quintessential field. The solution reduces to Schwarzschild spacetime when $c = 0$, and to other known solutions for specific values of $\omega$, such as $\omega = 1/3$ for radiation-like fields or $\omega = -1$ for a cosmological constant. This solution is particularly useful for exploring modifications to black hole geometries induced by exotic matter distributions. From Eq.~(\ref{eq21}), we conclude that the mass function is $m(r)=m_0 + c/2r^{3w}$ such that regularized geometry can be written as 
\begin{equation}
M(r) =\left(m_0 +\frac{c}{2r^{3w}}\right)\, \exp\left(-\frac{a^{n+1}}{r^{n+1}}\right).
\label{eq22}
\end{equation}
For $3w+1>1$, the geometry associated with this solution is asymptotically flat. For $3w+1 < 1$, cosmological horizons may appear depending on the values of the parameter of the equation of state. Figure \ref{fig3} depicts the behavior of the potential $V(r)$, where the curves associated with the regular solutions can be observed. 
By solving Eq.~(\ref{eq8}) for $B(r)$ associated with regularized Kiselev solution, we conclude that the associated energy density is given by
\begin{equation}
\rho(r) =e^{-\frac{a^2}{r^2}} r^{-5 - 3w} (8\pi)^{-1} \left[2 a^2 \left(c + 2 m_0 r^{3w}\right) - 3 c r^2 w\right].
\end{equation}
Now the Kiselev solution is regularized, and the energy density is modified, incorporating the influence of the exponential regularization factor controlled by the parameter $a$. In the regime where $a \ll r$, the first term inside the brackets can be neglected, so that the energy density reduces to $\rho(r)=-3cwr^{-3(w+1)}/8\pi$, in this case $\rho(r)$ is positive if $cw<0$. However, near $r = 0$, the energy density tends to zero, as expected due to the presence of the exponential factor.

\section{Final remarks}
In this work we proposed a method that offers an alternative to the singularity avoidance approach introduced by Simpson and Visser \cite{Simpson:2018tsi}, based on the coordinate transformation \( r \rightarrow \sqrt{r^{2}+a^{2}} \), and also to the proposal by Bronnikov \cite{Bronnikov:2024izh}, which applies a Bardeen-inspired modification directly to the metric. The prescription consists of replacing the mass function $m(r)$ by $M(r) = m(r)\exp(-a^{\,n+1}/r^{\,n+1})$, which guarantees that curvature invariants remain finite in the limit $r \to 0$ as long as $m(r)$ does not grow faster than the exponential function in Eq.~(\ref{eq11}). This simple modification of the mass function provides a unified language to treat different geometries and makes clear the analogy with regularization procedures in QFT, where divergences are softened by an exponential suppression at high energies.  

The analysis of geodesic completeness revealed that the choice of the parameter $n$ plays a decisive role in the global structure of the spacetime. For certain values of $n$, the metric can be extended to the region $r < 0$ in a smooth way, ensuring that geodesics can be extended for arbitrary affine parameter. This feature highlights that avoiding curvature divergences is not sufficient by itself, since geodesic completeness adds an independent criterion for regularity. We also observed that the horizon structure of the regularized solutions can be characterized in terms of the Lambert $W$ function, which allows a compact analytic expression for the radii of the inner and outer horizons. This result illustrates how the exponential cutoff not only removes singular behavior but also leads to new mathematical structures in the analysis of black holes.  

As an explicit application, we constructed a new regular version of the Kiselev solution, where the effects of quintessence are consistently incorporated into the exponential scheme. In this way, the method not only reproduces well-known geometries such as Schwarzschild or Reissner–Nordström, but also generates genuinely new spacetimes with different physical properties. In the case of the regularized Kiselev solution, the energy density acquires an exponential suppression at small $r$, ensuring that the geometry remains finite everywhere while still retaining the expected asymptotic behavior for large distances.  

Overall, the exponential cutoff approach provides a flexible tool to construct singularity-free black holes within classical general relativity. It establishes a clear parallel between the treatment of ultraviolet divergences in QFT and the resolution of curvature singularities in gravity. Future work could address possible extensions to rotating geometries. It would also be natural to investigate the thermodynamic stability and possible observational imprints of these regularized spacetimes.

\section{acknowledgments}
LCNS would like to thank Conselho Nacional de Desenvolvimento Científico e Tecnológico - Brazil (CNPq) for financial support under Research Project No. 443769/2024-9 and Research Fellowship No. 314815/2025-2.

\bibliography{references}

\end{document}